\pdfoutput=1
\RequirePackage[T1]{fontenc}
\RequirePackage{fix-cm}
\RequirePackage{lmodern}
\documentclass[conference]{IEEEtran}
\usepackage{cite}
\usepackage{amsmath,amssymb,amsfonts}
\usepackage[pass]{geometry}

\newcommand{\ket}[1]{|#1 \rangle}

\usepackage{stix2}

\newtheorem{theorem}{Theorem}

\newtheorem{remark}[theorem]{Remark}

\begin{document}

\title{Advance Sharing with Ogawa et al.'s Ramp Quantum Secret Sharing Scheme}

\author{\IEEEauthorblockN{Satoshi MASUMORI and Ryutaroh MATSUMOTO}
\IEEEauthorblockA{\textit{Dept.\  of Information and Communications Engineering} \\
\textit{Institute of Science Tokyo}\\
Tokyo, Japan \\
ORCID 0000-0002-5085-8879}}

\maketitle

\begin{abstract}
  The ramp quantum secret sharing proposed by Ogawa et al.\ has the
  highest possible coding rate given a threshold type access structure.
  On the other hand, in some quantum secret sharing schemes,
  it is known that some shares can be distributed to participants before
  a secret is given to the dealer.
  However, it is unclear whether some shares can be distributed before
  a secret is given in Ogawa et al.'s scheme.
  In this paper, we propose a method to distribute some shares
  before
  a secret is given in Ogawa et al.'s scheme,
  then determine a necessary and sufficient condition on sets of shares
  that can be distributed before a given secret.
\end{abstract}


\section{Introduction}
The study of quantum information processing gains
much attention recently.
One reason is that increase in the size of quantum computers
\cite{googleqc}.
The storage and communications of quantum information
are still difficult experimentally, but they will become easier and less expensive
in a future.
Classical secret sharing, in which secrets and shares are both classical information,
is nowadays used in practice, for example, in distributed storage systems.
In a distributed storage system \cite{attasena2017}, data are stored in multiple storages,
which increases the risk of data leakage. Secret sharing schemes decreases that risk
with the multiple storages.
It is likely that quantum secret sharing schemes play similar roles
in the future quantum information era.

In this paper, we consider to share a quantum secret by quantum shares.
We assume no interactions among the dealer and the participants
except distribution of shares from the dealer to the participants.
Such a problem formulation was initiated by Cleve, Gottesman and Lo
\cite{cleve99,gottesman00}, which is the most natural quantum
counterpart of the classical secret sharing
considered by Shamir \cite{shamir79} and Blakley \cite{blakley79}.

The coding rate is an important parameter in secret sharing schemes.
It is defined as the ratio of secret size to average of share sizes \cite{ogawa05}.
Another important property of secret sharing schemes
is the access structure, which consists of three families of qualified sets, forbidden sets,
and intermediate sets \cite{ogawa05}.
A set $S$ of shares is called qualified (resp.\  forbidden)
if $S$ allows reconstruction of the secret (resp.\ has no information about the secret).
A set $S$ of shares is called intermediate if it is neither qualified nor forbidden.
If a secret sharing scheme has no intermediate set, it is called perfect.
The coding rate cannot be greater than $1$ if it is perfect \cite{gottesman00}.
Ogawa et al.\ proposed ramp quantum secret sharing schemes,
which enable coding rates greater than $1$, at the cost of allowing intermediate sets.
In this paper, we consider $p$-dimensional quantum system $\mathcal{H}_p$
and call it a qudit, where $p$ is a prime number.
A $(k, L, n)$ quantum secret sharing scheme encodes
$L$-qudit quantum secret into $n$ shares each of which is 1 qudit in $\mathcal{H}_p$.
In a $(k, L, n)$ scheme, a set $S$ is qualified if $|S| \geq k$, forbidden if $|S| \leq k-L$,
and intermediate otherwise (i.e.\ $k-L+1 \leq |S| \leq k-1$).
Among $(k, L, n)$ schemes, Ogawa et al.'s one has the highest possible coding
rate $L$, and we focus on Ogawa et al.'s ramp quantum secret sharing.
The scheme by Cleve, Gottesman and Lo \cite{cleve99} is a special
case of Ogawa et al.'s corresponding to the case $L=1$.

It is sometimes convenient to distribute shares before a secret is given
to the dealer. One example of such situations was discussed in
\cite{miyajima22}, which considered sharing a classical secret
by quantum shares. Such distribution of shares before a given secret 
is named ``advance sharing'' \cite{miyajima22}.
Advance sharing is a trivial problem when both secret and shares are classical.
For example, encoding 1-bit secret $s$ into 2 shares $(x, s+x)$ provides a $(1,2)$
perfect secret sharing scheme and the first share $x$ is clearly advance shareable,
where $x$ is a random bit.
On the other hand, when shares are quantum, it is nontrivial to
realize advance sharing. In this paper, we focus on the case of quantum secrets.
The first quantum advance sharing was realized in Lie and Jeong
\cite{lie20}, which enabled advance sharing for some specific schemes.
On the other hand, any quantum error-correcting code and
its erasure decoding algorithm can be used as a quantum secret sharing
scheme \cite{cleve99,gottesman00}, and Shibata and Matsumoto \cite{shibata24}
clarified how to realize advance sharing with quantum secret sharing scheme
constructed from a $p$-adic quantum stabilizer code.

The ramp quantum secret sharing scheme by Ogawa et al.\ \cite{ogawa05}
has the highest coding rate as described before, but it is unclear how to
realize advance sharing with it. In this paper, we describe
Ogawa et al.'s scheme  \cite{ogawa05} as a stabilizer-based scheme,
and clarify which sets of shares are advance shareable,
based on the work in \cite{shibata24}.

This paper is organized as follows:
In Section \ref{sec2}, we review necessary contents from
\cite{ogawa05}, in addition to the quantum stabilizer codes \cite{matsumoto21}.
In Section \ref{sec3},
we describe Ogawa et al.'s scheme \cite{ogawa05} by a quantum stabilizer, which
allows direct application of results in \cite{shibata24}.
Then, by using a necessary and sufficient condition on sets of advance
shareable shares, we clarify which shares are advance shareable in \cite{ogawa05}.
A simple example is also given.
In Section \ref{sec4}, we give concluding remarks and future research agendas.

\section{Preliminaries}\label{sec2}
As defined before,
$\mathcal{H}_q$ denotes a $q$-dimensional quantum system
called a qudit, where $q$ is a prime \emph{power}.
Ogawa et al.\ assumed that shares belong to $\mathcal{H}_q$.
On the other hand,
Shibata and Matsumoto assumed that shares belong to $\mathcal{H}_q$
and that $q$ is a prime $p$ instead of a prime power.
In this paper,
we assume that shares belong to $\mathcal{H}_p$ for some prime $p$.
We start this section by reviewing Ogawa et al.'s scheme \cite{ogawa05}.

\subsection{Quantum Ramp Secret Sharing Scheme in \protect\cite{ogawa05}}
Let $\mathbf{F}_p$ be the finite field with $p$ elements,
and $\{ \ket{i} : i \in \mathbf{F}_p \}$ be an orthonormal basis
of $\mathcal{H}_p$.
For $\vec{s}=(s_1$, \ldots, $s_L) \in \mathbf{F}_p^L$
we define
\[
\ket{\vec{s}} = \ket{s_1} \otimes \cdots \otimes \ket{s_L} \in
\mathcal{H}_p^{\otimes L}.
\]
Let $n$ be the number of shares/participants, and we assume $n<p$.
Let $\alpha_1$, \ldots, $\alpha_n$ be distinct nonzero\footnote{%
Ogawa et al.\ \cite{ogawa05} allowed zero but it was a mistake.}
elements in $\mathbf{F}_p$.
We will review the construction of a $(k,L,n)$ ramp quantum scheme.
For $\vec{c} = (c_1$, \ldots, $c_k) \in \mathbf{F}_p^k$,
we define the polynomial $f_{\vec{c}}(x)$ by
\[
f_{\vec{c}}(x) = c_1 + c_2x+ \cdots + c_k x^{k-1}.
\]
A quantum secret $\ket{\vec{s}} \in \mathcal{H}^{\otimes L}$
is encoded to
\begin{equation}
  \frac{1}{\sqrt{p^{k-L}}}\sum_{\vec{c}\in D(\vec{s})} \ket{f_{\vec{c}}(\alpha_1)} \otimes \ket{f_{\vec{c}}(\alpha_2)} \otimes \cdots \otimes \ket{f_{\vec{c}}(\alpha_n)}, \label{eq10}
\end{equation}
where $D(\vec{s})$ is the set of vectors $\vec{c} \in \mathbf{F}_p^k$
whose leftmost $L$ components are the same as those in $\vec{s}$.

\subsection{Quantum Stabilizer Codes}
We define $p \times p$ unitary matrices
$X_p$ and $Z_\omega$
by $X_p\ket{i} = \ket{i+1 \bmod p}$ and $Z_\omega\ket{i} = \omega^{i} \ket{i}$,
where $\omega$ is a complex primitive $p$-th root of $1$.
Let $E_n = \{ \omega^i X_p^{x_1} Z_\omega^{y_1} \otimes \cdots
\otimes X_p^{x_n} Z_\omega^{y_n}$ : $i, x_1$, \ldots, $x_n$, $y_1$,
\ldots, $y_n \in \{0$, \ldots, $p-1\} \}$,
which can be viewed as a non-commutative group with matrix multiplication as
its group operation.
Let $S$ be a commutative subgroup of $E_n$, which is called a stabilizer.
Let $Q$ be a simultaneous eigenspace of all the matrices in $S$.
Then $Q$ is called a stabilizer code.
When $S$ is minimally generated by $k$ matrices (not counting a scalar multiple of the identity
matrix in $\mathcal{H}_p^{\otimes n}$, $Q$ encodes $k$-qudit message into
$n$-qudit codeword. Each codeword $\ket{\varphi} \in Q$ is an eigenvector
of every matrix in $S$.

\section{Advance Sharing with Ogawa et al.'s Ramp Quantum Secret Sharing Scheme}\label{sec3}
\subsection{Description of Ogawa et al.'s Ramp Quantum Secret Sharing Scheme by a Stabilizer}
In this section we give a description of Ogawa et al.'s scheme \cite{ogawa05}
by a stabilizer. Let
\begin{eqnarray*}
  C_1 &=& \{ (f_{\vec{c}}(\alpha_1), \ldots, f_{\vec{c}}(\alpha_n)) : \vec{c} \in \mathbf{F}_p^k \},\\
  C_2 &=& \{ (f_{\vec{c}}(\alpha_1), \ldots, f_{\vec{c}}(\alpha_n)) : \vec{c} \in D(\vec{0}) \}.
\end{eqnarray*}

\begin{theorem}\label{thm1}
  Let $S$ be the stabilizer of $E_n$ generated by
  $G_1 = \{ Z_\omega^{x_1} \otimes \cdots \otimes Z_\omega^{x_n}$ :
  $(x_1$, \ldots, $x_n) \in C_1^\perp \}$
  and $G_2  = \{ X_p^{x_1} \otimes \cdots \otimes X_p^{x_n}$ :
  $(x_1$, \ldots, $x_n) \in C_2 \}$.
  Let $Q$ be a complex linear space generated by
  \begin{equation}
    \ket{\varphi} = \frac{1}{\sqrt{p^{k-L}}}\sum_{\vec{c}\in D(\vec{s})} \ket{f_{\vec{c}}(\alpha_1)} \otimes \ket{f_{\vec{c}}(\alpha_2)} \otimes \cdots \otimes \ket{f_{\vec{c}}(\alpha_n)},
      \label{eq1}
  \end{equation}
  for all $\vec{s} \in \mathbf{F}_p^L$.
  Then $Q$ is the stabilizer code defined by $S$.
\end{theorem}
\begin{IEEEproof}
  Firstly we will prove that any quantum state in $Q$ is contained in the
  stabilizer code defined by $S$. To do so,
  it is enough to verify that every vector $\ket{\varphi} \in Q$
  belongs to the same eigenvalue of an arbitrarily fixed $M \in G_1 \cup G_2$.
  Without loss of generality we may assume that $\ket{\varphi}$ is written as
  Eq.\ (\ref{eq1})    for some $\vec{s} \in \mathbf{F}_p^L$.

    Let assume $M \in G_1$.
  Then $M$ can be written as
  $Z_\omega^{x_1} \otimes \cdots \otimes Z_\omega^{x_n}$ for some
  $\vec{x} = (x_1$, \ldots, $x_n) \in C_1^\perp$.
    Then we have
    \begin{eqnarray*}
    &&Z_\omega^{x_1} \otimes \cdots \otimes Z_\omega^{x_n} \ket{f_{\vec{c}}(\alpha_1)} \otimes \ket{f_{\vec{c}}(\alpha_2)} \otimes \cdots \otimes \ket{f_{\vec{c}}(\alpha_n)}
    \\&=& \omega^\ell \ket{f_{\vec{c}}(\alpha_1)} \otimes \ket{f_{\vec{c}}(\alpha_2)} \otimes \cdots \otimes \ket{f_{\vec{c}}(\alpha_n)},
    \end{eqnarray*}
    where $\ell$ is the Euclidean inner product between $\vec{x}$
    and $(f_{\vec{c}}(\alpha_1)$, \ldots, $f_{\vec{c}}(\alpha_n))$,
    which is equal to zero as $\vec{x} \in C_1^\perp$.
    Therefore, we see
    \[
    Z_\omega^{x_1} \otimes \cdots \otimes Z_\omega^{x_n} \ket{\varphi} = \ket{\varphi}
    \]
    independently of the value of $\vec{s} \in \mathbf{F}_p^L$.

    Let assume $M \in G_2$.
  Then $M$ can be written as
  $X_p^{x_1} \otimes \cdots \otimes X_p^{x_n}$ for some
  $\vec{x} = (x_1$, \ldots, $x_n) \in C_2$.
  The multiplication by $X_p^{x_1} \otimes \cdots \otimes X_p^{x_n}$
  just permutes indices $\vec{c}$ of terms in Eq.\ (\ref{eq1}).
  Therefore $X_p^{x_1} \otimes \cdots \otimes X_p^{x_n} \ket{\varphi}= \ket{\varphi}$.

  By the argument up to now, we see that for any $M\in S$ we have
  $M\ket{\varphi}=\ket{\varphi}$.
  Since any vector in $Q$ is a linear combination of vectors $\ket{\varphi}$
  of Eq.\ (\ref{eq1}), every vector in $Q$ belongs to eigenvalue $1$ of every $M \in S$.
  We have verified that $Q$ is contained in the stabilizer defined by $S$.

  Secondly, the stabilizer defined by $S$ encodes $L$ qudits in $\mathcal{H}_p^{\otimes L}$.
  On the other hand, since $p^L$ orthonormal vectors in Eq.\ (\ref{eq1})
  generates $Q$, $\dim Q=p^L$, which implies that
  $Q$ is equal to the stabilizer code defined by $S$.
\end{IEEEproof}

Shibata and Matsumoto \cite{shibata24}
showed how to realize advance sharing
for any stabilizer-based quantum secret sharing
by modifying encoder of secrets into shares while retaining the coding rate
and the access structure.
Now it is clear that some shares are advance shareable in Ogawa et al.'s scheme.
The remaining task is to clarify which shares are advance shareable.

\subsection{Advance Shareable Sets in Ogawa et al.'s Scheme}
It is well-known that a stabilizer $S$ can be described by
a self-orthogonal linear code $C \subset \mathbf{F}_p^{2n}$, see, e.g.\  \cite{matsumoto21}.
Shibata and Matsumoto \cite{shibata24} showed a necessary and sufficient
condition on which sets of shares are advance shareable by using their
proposal.

For a subset $J \subset \{1$, \ldots, $n\}$,
we define the shortened code $C_{J}$ with respect to $J$
as
\[
\{ (a_1, a_2, \ldots, a_n | b_1, b_2, \ldots, b_n) \in C :
a_i = b_i =0, \forall i \in J \}.
\]
A set $J$ is advance shareable by the proposal in \cite{shibata24}
if and only if
\begin{equation}
  \dim C_{J} = \dim C - 2|J|, \label{eq2}
\end{equation}
where $C$ defines the underlying stabilizer of the quantum secret sharing scheme.

\begin{theorem}\label{thm2}
  In Ogawa et al.'s secret sharing scheme,
  a set $J$ of shares is advance shareable if and only if
  $|J| \leq k-L$.
\end{theorem}
\begin{IEEEproof}
  As stated in \cite[Lemma 2]{ogawa05}, we have
  \begin{equation}
    n=2k-L. \label{eq3}
  \end{equation}
  In Ogawa et al.'s scheme, $C = C_1^\perp \times C_2$ by Theorem \ref{thm1}.
  Both $C_1^\perp$ and $C_2$ are $[n,k-L, n-k+L+1]$ generalized Reed-Solomon codes, and
  their Euclidean duals $C_1$ and $C_2^\perp$ are $[n,n-k, k-L+1]$ generalized Reed-Solomon codes \cite{pless98}
  by Eq.\ (\ref{eq3}). If $|J| \leq k-L$ then Eq.\ (\ref{eq2}) holds
  by \cite[Lemma 10.1]{pless98}.
 
  Let $(C_2)_J$ be the shortened code with respect to $J$ and $C_2^J$
  the punctured code with respect to $J$, see, e.g., \cite{pless98} for a definition.
  Then we have $\dim (C_1^\perp)_J + \dim C_1^J = n-|J|$
  and $\dim (C_2)_J + \dim (C_2^\perp)^J = n-|J|$ \cite[Lemma 10.1]{pless98}.
  Since $\dim C_1 = \dim C_2^\perp = n-k = k-L$, $\dim C_1^J \leq k-L$
  and $\dim (C_2^\perp)^J \leq k-L$.
  Therefore, $\dim (C_1^\perp)_J \geq n-(k-L)$ and
  $\dim (C_2)_J \geq n-(k-L)$.
  This means that if $|J| \geq k-L+1$ then Eq.\ (\ref{eq2}) cannot hold.
\end{IEEEproof}

\begin{remark}
  By Theorem \ref{thm2} and \cite{ogawa05},
  in Ogawa et al.'s scheme, a set of shares is advance sharable if and only if
  it is forbidden. This equivalence does not hold in general. In general, an advance shareable
  set is always forbidden, but the converse is not always true, see examples in
  \cite{shibata24}.
\end{remark}

\subsection{Simple Example}
Let $p=5$, $n=4$, $k=3$, $L=2$, and
$\alpha_i = i \in \mathbf{F}_5$.
The standard encoding procedure in Eq.\ (\ref{eq10})
encodes a quantum secret $\ket{\vec{s}}$ for $\vec{s} =(s_1, s_2) \in \mathbf{F}_5^2$
into
\begin{eqnarray}
  &&\frac{1}{\sqrt{5}}\sum_{\vec{c}\in D(\vec{s})} \ket{f_{\vec{c}}(1)} \otimes \ket{f_{\vec{c}}(\alpha_2)} \otimes \ket{f_{\vec{c}}(\alpha_3)} \otimes \ket{f_{\vec{c}}(\alpha_4)} \nonumber\\
    &=&
    \frac{1}{\sqrt{5}}\bigl(
    \ket{s_1+s_2+1, s_1+2s_2+4, s_1+3s_2+9, s_1+4s_2+16} \nonumber\\
&&+  \ket{s_1+s_2+2\cdot 1, s_1+2s_2+2\cdot 4, s_1+3s_2+2\cdot 9, s_1+4s_2+2\cdot 16}\nonumber\\
&&+  \ket{s_1+s_2+3\cdot 1, s_1+2s_2+3\cdot 4, s_1+3s_2+3\cdot 9, s_1+4s_2+3\cdot 16}\nonumber\\
    &&+  \ket{s_1+s_2+4\cdot 1, s_1+2s_2+4\cdot 4, s_1+3s_2+4\cdot 9, s_1+4s_2+4\cdot 16}\bigr). \nonumber\\
    &&\label{eq4}
\end{eqnarray}
As every qudit in Eq.\ (\ref{eq4}) depends on the quantum secret $\ket{\vec{s}}$,
from the encoding procedure in \cite{ogawa05}, it is completely unclear
(to the authors) how one share is advance shareable.
By Theorem \ref{thm2} any set $J$ of shares is advance shareable if $|J| = 1$.
By applying \cite{shibata24} one can make the set $J=\{1\}$,
$\{2\}$, $\{3\}$, or $\{4\}$ advance shareable,
see \cite{shibata24} for the details.

\section{Concluding Remarks}\label{sec4}
In this paper, we showed how to make shares in Ogawa et al.'s ramp
quantum secret sharing scheme advance shareable,
and also clarified which sets are advance shareable.
On the other hand, in \cite{ogawa05},
the dimension of a share is assumed to be a prime power,
but this paper imposes a more strict requirement being a prime $p$.
In a forthcoming journal paper version of this conference paper,
we plan to remove this extra requirement.

\section*{Acknowledgment}
This research is in part supported by the JSPS grant No.\ 23K10980.



\end{document}